\documentstyle[epsfig,12pt]{article}  
\newcommand{\bce}{\begin{center}}
\newcommand{\ece}{\end{center}}
\newcommand{\beq}{\begin{equation}}
\newcommand{\eeq}{\end{equation}}
\newcommand{\bea}{\vspace{0.25cm}\begin{eqnarray}}
\newcommand{\eea}{\end{eqnarray}}

\newcommand{\ba}{\begin{array}}
\newcommand{\ea}{\end{array}}

\newcommand{\r}{\mbox{{\boldmath
$\rho$}}}

\newcommand{\singlespace}{
\renewcommand{\baselinestretch}{1}\large\normalsize}
\newcommand{\doublespace}{
\renewcommand{\baselinestretch}{1.6}\large\normalsize}

\def\lsim{\mathrel{\rlap{\lower4pt\hbox{\hskip1pt$\sim$}}
    \raise1pt\hbox{$<$}}}         
\def\gsim{\mathrel{\rlap{\lower4pt\hbox{\hskip1pt$\sim$}}
    \raise1pt\hbox{$>$}}}         

    \def\beq{\begin{equation}}
    \def\endeq{\end{equation}}
    \def\bea{\begin{eqnarray}}
    \def\arr{\begin{eqnarray}}
    \def\eea{\end{eqnarray}}

\def\q2{$Q^{2}$}
\def\s2{2$S$}

\begin{document}
\doublespace

\begin{center}

  {\Large\bf

Description of the CERN SPS data on the Landau-Pomeranchuk-Migdal effect 
for photon bremsstrahlung in quantum regime
  }

\vspace{1.0cm}

  {\large B.G. Zakharov  }
\vspace{0.5cm}

{
\singlespace
L.~D.~Landau Institute for Theoretical
Physics,
GSP-1, 117940,\\ ul. Kosygina 2, 117334 Moscow,
Russia
\\}
\vspace{.5cm}
{\bf
Abstract}
\end{center}
{
We analyze within the light-cone path integral
approach \cite{ZLPM1} the recent CERN SPS data \cite{SPS1} on
the Landau-Pomeranchuk-Migdal effect for photon bremsstrahlung
in quantum regime from 149, 207, and 287 GeV electrons.
Calculations have been carried out treating accurately the 
Coulomb effects and including the inelastic processes.
Comparison with experiment is performed 
accounting for multi-photon emission. Our results 
are in good agreement with the data.}\\

\vspace{0.5cm}
\noindent{\bf 1.} 
Recently much attention has been attracted to suppression of 
radiation processes in media due to multiple scattering (the 
Landau-Pomeranchuk-Migdal (LPM) effect \cite{LP,Migdal})
both in QED \cite{BlanD,ZLPM1,B2,ZLPM2,Blan,BK1} and 
QCD \cite{BDPS,BDMPS,ZLPM1,ZLPM5} (for recent reviews,  see \cite{Klein,BSZ}).
In QED new interest in the LPM effect stems from the first accurate
measurements of the LPM suppression by the SLAC
E-146 collaboration \cite{SL2} for photon bremsstrahlung
from 8 and 25 GeV electrons. 
A detailed analysis of these high statistics data requires more accurate
calculations than that perform by Migdal \cite{Migdal} 
within the Fokker-Planck approximation.
The interest in the LPM effect in QCD is connected with the study of parton
energy loss in a hot QCD matter which can be produced in high energy 
$AA$-collisions at RHIC and LHC.

In \cite{ZLPM1} (see also \cite{ZLPM2,ZLPM3,ZLPM4}) we have developed a new 
approach to the LPM effect, which we call the light-cone path integral
(LCPI) approach. This formalism is applicable for radiation processes 
in both QED and QCD.
It is based on the path integral treatment of multiple scattering
in the transverse coordinate representation developed in \cite{A2eBGZ}.
In \cite{ZLPM6} we have analyzed within the LCPI approach the SLAC data 
\cite{SL2}, and obtained excellent agreement with
the data. For the data of Ref. \cite{SL2} the photon fractional momentum
is small $x\lsim 0.06$ ($x=k/E$, $k$ is the photon momentum, 
$E$ is the beam energy), and the LPM suppression becomes visible 
for $x\lsim 0.02$ where quantum recoil effects are unimportant.
In the present paper we analyze within our
approach new data on the LPM effect obtained in the CERN SPS experiment
\cite{SPS1} for bremsstrahlung from 147, 207, and 287 GeV electrons. 
It is of clear interest because the SPS data allow one, for the first time,
to compare theory with experiment in the kinematic region
where quantum effects come into play. The kinematic range covered by
the SPS data \cite{SPS1} is $0.01\lsim x<1$, and
the onset of the LPM effect occurs for $x\sim 0.1$ where quantum effects
are important.

\noindent{\bf 2.} In the LCPI approach the cross section of an induced $a\rightarrow bc$ 
transition is expressed 
through the solution of a two-dimensional
Schr\"odinger equation with an imaginary potential
which is proportional to the cross section of interaction
of $\bar{a}bc$ system  with a medium constituent. For
$e\rightarrow \gamma e$ 
transition the corresponding cross section is 
the dipole cross section for scattering of $e^{+}e^{-}$ pair off an
atom, $\sigma(\rho)$ (here $\rho$ is the transverse distance between 
electron and positron), and the Hamiltonian reads
\beq
{
H}=-\frac{1}{2\mu(x)}
\frac{\partial^{2}}{\partial \r^{2}}
+v(\r,z)\,,
\label{eq:2.1}
\eeq
\beq
v(\r,z)=-i\frac{n(z)\sigma(|\r
|x)}{2}\,,
\label{eq:2.2}
\eeq
where the Schr\"odinger mass
is $\mu(x)=Ex(1-x)$, and $n(z)$ is the number density of the target.
The longitudinal coordinate $z$ in (\ref{eq:2.1}), (\ref{eq:2.2}) 
plays the role of time. 
The probability distribution of one-photon emission, $dP/dx$,
is given by \cite{ZLPM1}
\beq
\frac{d P}{d
x}=2\mbox{Re}
\int\limits_{-\infty}^{\infty} d
z_{1}
\int\limits_{z_{1}}^{\infty}d
z_{2}
\exp\left(-\frac{i\Delta z}{L_{f}}\right)
\hat{g}
\left[{\cal
K}(\r_{2},z_{2}|\r_{1},z_{1})
-{\cal
K}_{v}(\r_{2},z_{2}|\r_{1},z_{1})
\right]\left.\right|_{\r_{1}=\r_{2}=0}\,.
\label{eq:2.3}
\eeq
Here $\Delta z=z_{2}-z_{1}$, ${\cal K}$ is the Green's function for the
Hamiltonian (\ref{eq:2.1}), ${\cal K}_{v}$ is the vacuum Green's
function for $v(\r,z)=0$,
$
L_{f}=
{2E(1-x)}/{m_{e}^{2}x}\,,
$
$\hat{g}$
is the vertex operator accumulating spin effects (for  its explicit form
see \cite{ZLPM1}).

The dipole cross section can be written as $\sigma(\rho)=C(\rho)\rho^{2}$, 
where 
\beq
C(\rho)=Z^{2}C_{el}(\rho)+Z
C_{in}(\rho)\,.
\label{eq:2.4}
\eeq
Here the terms $\propto Z^{2}$ and $\propto
Z$ correspond to elastic and inelastic interactions of $e^{+}e^{-}$ pair 
with an atom. In the region $\rho\lsim 1/m_e$, which is important for 
evaluation of the radiation rate (see below),
$C_{el}(\rho)$ can be parametrized as
\bea
C_{el}(\rho)=
4\pi\alpha^{2}
\left[\log\left(\frac{2a_{el}}{\rho}\right)+\frac{(1-2\gamma)}{2}
-f(Z\alpha)\right]\,,
\label{eq:2.5}
\eea
$$
f(y)=y^{2}\sum\limits_{n=1}^{\infty}
\frac{1}{n(n^{2}+y^{2})}\,\,,
$$
where $\alpha=1/137$, $\gamma=0.577$ is Euler's constant. The $C_{in}(\rho)$ can be 
written in a similar form (with $a_{el}$ replaced by $a_{in}$) but without 
$f(Z\alpha)$. 
The parametrization of $C_{el}(\rho)$ corresponds to scattering of
$e^{+}e^{-}$ pair on the atomic potential $\phi(r)=(Ze/4\pi r)\exp(-r/a_{el})$.
The first two terms in the square brackets on the right-hand side of 
(\ref{eq:2.5}) 
stem from the Born approximation, while the last one
gives the Coulomb correction due to multi-photon exchanges. 
To account for the finite size of nucleus one should replace on the 
right-hand side of (\ref{eq:2.5}) $\rho$ by $R_{A}$ for $\rho\lsim R_{A}$ 
(here $R_{A}$ is the nucleus radius). 

The parameters $a_{el}$ and $a_{in}$ can be adjusted by comparing
the Bethe-Heitler cross section calculated via the dipole cross section
\cite{ZLPM2}
\beq
\frac{d \sigma^{BH}}{d
x}=
\int
d\r\,
|\Psi_{e}^{e\gamma}(x,\r)|^{2}
\sigma(\rho
x)\,\,
\label{eq:2.8}
\eeq
(here $\Psi_{e}^{e\gamma}(x,\r)$ is the light-cone
wave function for the $e\rightarrow e\gamma$ transition)
with that
obtained within the standard approach using the Thomas-Fermi-Molier model
\cite{Tsai}.
This gives $a_{el}=0.81r_{B}Z^{-1/3}$ and $a_{in}=5.3r_{B}Z^{-2/3}$ 
\cite{ZLPM6}.

Treating the potential (\ref{eq:2.2}) as a perturbation one can represent 
the spectrum (\ref{eq:2.3}) as
\beq
\frac{d P}{d
x}=
\frac{d P^{BH}}{d x}+\frac{d P^{abs}}{d
x}\,,
\label{eq:2.9}
\eeq
where the first term
\beq
\frac{d P^{BH}}{d
x}=nL
\frac{d \sigma^{BH}}{d
x}\,
\label{eq:2.10}
\eeq
is the Bethe-Heitler spectrum ($L$ is the target thickness), 
and the second one is the absorptive correction
responsible for LPM suppression  \cite{ZLPM2}.
We use this representation for the numerical calculations.
The explicit form of the absorptive term and light-cone
wave function entering (\ref{eq:2.8}) can be found in \cite{ZLPM2}. 

An important characteristic of the LPM effect is the in-medium
coherence (formation) length of photon emission 
(we refer to it as $L_{f}^{eff}$). In the sense of the representation
(\ref{eq:2.3}) $L_{f}^{eff}$ is simply the typical $\Delta z$. The 
$L_{f}^{eff}$ is related to the dominating  $\rho$-scale, 
$\rho_{eff}$, via the diffusion relation 
$\rho_{eff}\sim \sqrt{2L_{f}^{eff}/\mu(x)}$.
From it follows that for small LPM suppression (when $L_{f}^{eff}\sim L_{f}$) 
$x\rho_{eff}\sim 1/m_{e}$. 
For strong LPM effect $L_{f}^{eff}$ and $\rho_{eff}$ are determined by the 
interplay of the diffusion and absorption effects.
From the diffusion relation and condition
$
nL_{f}^{eff}\sigma(x\rho_{eff})/2\sim 1\,,
$
saying that absorption for the $e^{+}e^{-}\gamma$ system becomes strong
at $\Delta z\sim L_{f}^{eff}$, one obtains
$
x\rho_{eff}\sim 1/{m_{e}}\sqrt{\eta}
$, and
$
L_{f}^{eff}\sim L_{f}/\eta
$,
where
$
\eta=2[nE(1-x)C(1/m_{e})/xm_{e}^{4}]^{1/2}
$. 
Thus one sees that, as was said above, the spectrum is only sensitive 
to the behavior of $C(\rho)$ at $\rho\lsim 1/m_{e}$.

Note that the Fokker-Planck approximation in momentum representation 
used in  Migdal's analysis \cite{Migdal}
corresponds to the replacement of $C(\rho)$ by $C(\rho_{eff})$.
Then the spectrum can be written via 
the oscillator Green's function. This approximation works well for strong
LPM effect ($\eta\gg 1$) when the logarithm in (\ref{eq:2.5}) 
is much larger than unity. For the SPS conditions $\eta\lsim 3$ and
the logarithm in (\ref{eq:2.5}) is $\sim 4-5$. In such a regime 
neglecting the logarithmic  dependence of $C(\rho)$ may give   
errors about 10-20\%.

\noindent{\bf 3.}
In \cite{SPS1} the experimenters used Ir target with thickness 
$L=4.36\%X_{0}=0.128$ mm, here $X_{0}$ is the radiation length,
for Ir $X_{0}\approx 2.94$ mm \cite{Tsai}.
The spectra in the radiated energy were measured for $k>2$ GeV.
Similarly to the SLAC data, the SPS spectra include multi-photon emission.
Using the given above estimate for $L_{f}^{eff}$ one can see that for the SPS
conditions $L_{f}^{eff}/L$ does not exceed $\sim 0.02-0.04$. 
Smallness of this quantity  allows one to evaluate
multi-photon emission in the probabilistic approach. 

We write the spectrum in the radiated energy (we refer to it as $dN/dx$) as
\beq
\frac{dN(E,x)}{dx}=K(E,x)\frac{dP(E,x)}{dx}\,,
\label{eq:3.1}
\eeq
where $K(E,x)$ accounts for multi-photon effects.
For the SPS kinematic domain $N$-photon emission with $N>2$ can be 
neglected.  Then in the probabilistic treatment of the radiation of one and 
two photons one can obtain for $K(E,x)$\footnote{
This is a generalization to arbitrary $x$ of the 
low-$x$ $K$-factor derived in \cite{ZLPM6}.}
\beq
K(E,x)=\exp{\left[-\int
\limits_{x}^{1}dy\frac{dP(E,y)}{dy}\right]}\cdot
\left[1+\frac{1}{2}F(E,x,\delta)\right]\,,
\label{eq:3.3}
\eeq
\bea
F(E,x,\delta)=
\int\limits_{\delta}^{\delta/(1-x)}dy\frac{dP(E,y)}{dy}+
\int\limits_{\delta}^{1}dy
\left[\frac{dP(E,y)}{dy}-\frac{dP(E(1-x),y)}{dy}\right]\nonumber\\
-\int\limits_{\delta}^{x-\delta}dx_{1}\left[
\frac{dP(E,x_{1})}{dx_{1}}+\frac{dP(E,x_{2})}{dx_{2}}\right.
\nonumber\\
-\left.
\frac{1}{1-x_{1}}\frac{dP(E,x_{1})}{dx_{1}}
\frac{dP(E(1-x_{1}),y_{2})}{dy_{2}}
\left(
\frac{dP(E,x)}{dx}
\right)^{-1}\right]
\,,
\label{eq:3.5}
\eea
where
$x_{2}=x-x_{1}$, $y_{2}=x_{2}/(1-x_{1})$, $\delta=k_{min}/E$. 
Although the expression (\ref{eq:3.5}) is infrared finite, 
we indicated explicitly
dependence on the low energy cutoff $k_{min}$.
In our numerical calculations we take $k_{min}=0.1$ GeV.
The results are practically insensitive to variation of this quantity.
The inaccuracy of the above expression for the $K$-factor at the SPS 
conditions is $\lsim 0.5$\%. This is clearly
not worse than the accuracy of the probabilistic approach by itself.

\noindent{\bf 4.} 
In Fig.~1 we plotted the SPS data and 
our results obtained with (solid line) and without (dashed line) LPM 
suppression. In \cite{SPS1} the spectra in the radiated energy were
presented in logarithmic bins (25 per decade). It corresponds to
$rdN/d\log{(k)}$ with $r=2\tanh{(\Delta/2)}$, $\Delta=\log{(10)}/25$. 
This form is used in Fig.~1.
Besides the multi-photon emission we have taken into account
photon absorption multiplying the theoretical spectra by 
$\langle K_{abs}\rangle\approx1-L/2\lambda_{ph}$,
where $\lambda_{ph}$ is the photon attenuation
length. This decreases the spectra by $\lsim 1.5$\%.
The theoretical spectra were multiplied by normalization coefficients,
$C_{norm}$, that were adjusted at each energy to minimize $\chi^{2}$.
We obtained the values 
$C_{norm}=0.994\pm 0.006$ ($\chi^{2}/N=1.08$), 
$0.982\pm 0.005$ ($\chi^{2}/N=1.78$),
and $0.944\pm 0.004$ ($\chi^{2}/N=1.43$) for $E=287,$ 207, and 149 GeV, 
respectively.
One can see that the theoretical spectra (with LPM suppression) 
are in good agreement with the data.

\noindent{\bf 5.} To summarize, we have analyzed the recent SPS data 
\cite{SPS1} on the LPM effect for bremsstrahlung from 149, 207, and 287 
GeV electrons. The calculations have been performed within the light-cone path
integral formalism \cite{ZLPM1,ZLPM2}. 
We treat accurately the Coulomb
effects and include the inelastic processes.
The comparison with experiment is performed 
accounting for multi-photon emission and photon absorption.
Our results  are in good agreement with the data
in the full studied kinematic range which includes 
the region of the onset of the LPM effect where
the photon fractional momentum $x\sim 0.1$, and quantum
recoil effects are important.\\
\phantom\\

I am grateful to Ulf-G.~Mei{\ss}ner and J.~Speth for the hospitality at 
FZJ, J\"ulich, where
this work was completed. 
I would also like to thank U.I.~Uggerh{\o}j for sending the files of
the experimental data and comments about some aspects of the SPS
experiment.
This work was partially supported by the grant
DFG 436RUS17/72/03.

\newpage
\begin{figure}[h]
\vspace*{-4cm}
\hspace*{-2.5cm}\epsfig{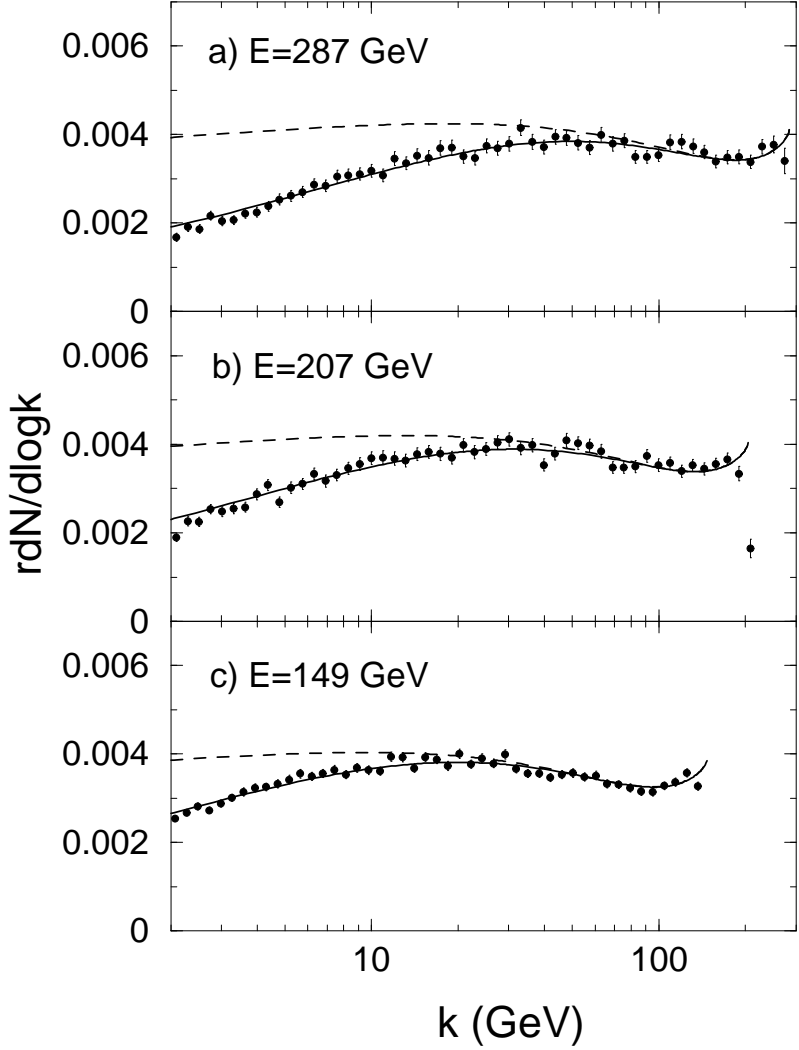}
\vspace*{-9cm}
\caption{
Comparison of the spectra in the radiated energy from the SPS experiment 
\cite{SPS1} with our calculations for bremsstrahlung from 
287~(a), 207~(b), and 149~(c) GeV electrons
on $4.36\%\,X_{0}$ Ir target.
The solid line shows our results
with LPM suppression. The dashed line shows
the Bethe-Heitler spectrum. The multi-photon emission and photon 
absorption are taken into account. The Bethe-Heitler spectra are 
multiplied by the normalization coefficients obtained 
for the spectra with LPM suppression.
}
\label{f1}
\end{figure}

\end{document}